\documentclass[5p,times]{elsarticle}

\PassOptionsToPackage{hyphens}{url}\usepackage{hyperref}
\usepackage{xspace}
\newcommand{\Purim}{\textsc{Purim}\xspace}
\newcommand{\PurimLong}{Purified by Unified Resampling of Infected Multitudes\xspace}

\usepackage{graphicx}
\usepackage{amsmath,amssymb}
\usepackage{ulem,xcolor}           
\usepackage{lipsum}                
\usepackage{enumitem}              
\setlist[enumerate]{itemsep=0mm}   
\graphicspath{{./HiResFig/}{./Fig/}{./}} 
\biboptions{sort&compress}
\journal{International Journal of Infectious Disease}  

\begin{document}

\begin{frontmatter}

\title{\textsc{Purim}: a rapid  method  with reduced cost for massive detection of CoVid-19}

      \author[]{Benjamin Isac Fargion}
      \author[sap,mifp]{Daniele Fargion\fnref{cor}} \ead{daniele.fargion@uniroma1.it}
      \author[RU]{{Pier Giorgio} {De Sanctis Lucentini}\fnref{PgDSLORCiD}}
      \author[DIAEE]{Emanuele Habib}

      \fntext[cor]{Corresponding author, ORCiD:0000-0003-3146-3932}
      \fntext[PgDSLORCiD]{ORCiD:0000-0001-7503-2064}

      \address[sap]{Physics Department \& INFN Rome1, Rome University 1, P.le A. Moro 2, 00185, Rome, Italy}
      \address[mifp]{Mediterranean Institute of Fundamental Physics, Via Appia Nuova 31, 00040 Marino (Rome), Italy}
      \address[RU]{Physics Department, Gubkin Russian State University (National Research
      University), 65 Leninsky Prospekt, Moscow, 119991, Russian Federation}
      \address[DIAEE]{DIAEE, Department Of Astronautical, Electrical And Energy Engineering, University of Rome "La Sapienza",  Via Eudossiana 18, Rome, Italy}




\begin{abstract}

The CoVid-19 is spreading pandemically all over the world. A rapid defeat of the pandemic requires carrying out on the population a mass screening, able to separate positive from negative cases. Such a cleaning will free a flow of productive population. The current rate and cost of testing, performed with the common PCR (polymerase chain reaction) method and with the available resources, is forcing a selection of the subjects to be tested. Indeed, each one must be examined individually at the cost of precious time. Moreover, the exclusion of potentially positive individuals from screening induces health risks, a broad slowdown in the effort to curb the viral spread, and the consequent mortality rates. We present a new procedure, the Purified by Unified Resampling of Infected Multitudes, in short Purim, able to untangle any massive candidate sample with inexpensive screening, through the cross-correlated analysis of the joint speciments. This procedure can reveal and detect most negative patients and in most cases discover the identity of the few positives already in the first or few secondary tests. We investigate the the two-dimensional correlation case in function of the infection probability. The multi-dimensional topology, the scaled Purim procedure are also considered. Extensive Purim tests may measure and weight the degree of epidemic: their outcome may identify focal regions in the early stages. Assuming hundreds or thousand subjects, the saving both in time and in cost will be remarkable. Purim may be able to filter scheduled flights, scholar acceptance, popular international event participants. The optimal extension of Purim outcome is growing as the inverse of the epidemia expansion. Therefore, the earlier, the better.

\end{abstract}
\end{frontmatter}


\section{Introduction}

 The CoVid-19 hidden pandemia, born recently \cite{huang2020clinical,bianconi2020oswald,cinelli2020covid, Gralinski2020, doi:10.1056/NEJMoa2001017,Zhou_2020} requires urgent detection for larger group sample, to disentangle positive and negative cases. The urgency of the screening  of the population cannot face easily the exponential infection growth. Any test examined individually, may take precious time, causing bottlenecks in the waiting list and reducing our ability to win the virus  exponential diffusion growth. 
 
 Because the possible geographical relation to the climate, \cite{Araujo2020.03.12.20034728},  selection test among countries and continents might be request for seasons and even for years. The mobility among nation depends on ability to recognize quickly the health  state of travelers. At the present time the rapid tests, as the ones based on the saliva (gingival exudate)\cite{to2020consistent} probe may offer, one by one, a validation.
  There is no other known  cheap and fast methods to test a large population at once.  Here we suggest a joint and unified unique test or a mixed  sub-groups test array (for instance by gingival exudate),  united into several independent ones, whose cross checked results  might discover in an rapid and economic way the main infected cases. The mathematical procedure might be exact or it might just lead  to an overestimate of the infected cases. Therefore, because of overestimation one could require  a few additional individual tests with marginal cost.
      
  The \PurimLong, \Purim\footnote{Let us remind that the word Purim, in Hebrews, means "fate"; the name of the Queen, Ester, means,   "hidden", as the Covid-19 pandemia behavior: the proposal was indeed  conceived recently, on the Purim 5780. During Purim, like Carnival, all  wear a mask ; the fate of all the people is turn suddenly from the disgrace to the life.}, may become the key strategy to separate, to trace the virus diffusion leading  to a sooner  suppression.\\
  Moreover, the screening procedure of a large population to avoid the spreading of the infection should detect any suspect case. The highlighting of suspect cases that are not actually positive to nCoV would increase the cost of the screening but won't reduce its effectiveness.
Most virus tests are individual, often testing in a blind and costly way wide numbers of the population. We believe that such single individual test politics in present pandemic stage is time consuming, costly and unable to face the fast virus growth and the largest population screening. Moreover, the foreseen  return of secondary waves of infection might require iterative or permanent filter of travelers from abroad. Quick and massive tests are needed. The procedure remind somehow the procedure to disentangle a scrambled card deck that is not well  shuffled \cite{fargion2014howwhen}.

\begin{figure}[!htb]
\begin{center}
\includegraphics[width=0.90\columnwidth]{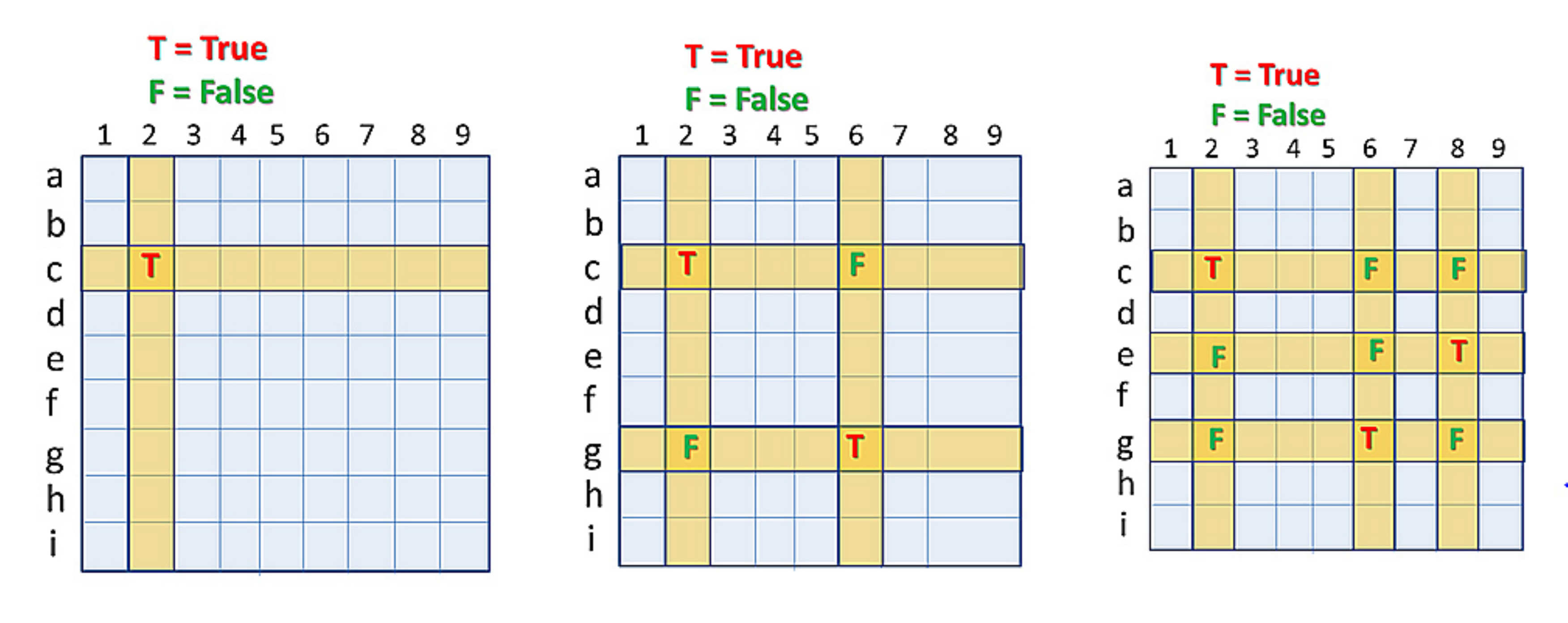}
\caption{\textsc{CrossCheck} The different cross-checked situations for a sample of $N=9$,  $N^2 = 81$ subjects, where one, or two or three different infected True positives candidates are present.  The sub-groups are just made by the horizontal and by the vertical column array $=2\cdot N=18$ sample of patients. Any sub-group test (composed by $N=9$ individuals each) is mixed and united with the all $N=9$ individual saliva. Their $2\cdot N$ array results are later cross-checked. False positive cases might also arise as a "mirror or shadows" effect. In a first case on the left, the single positive case may be immediately identify as an unique True positive case. Remaining candidates are negative. Al least $((81-(18+1)/81)= 79\%$ doubled checked negative. In the second and in the third arrays the presence of two or three True positive infect subjects produce more False positive  born by   the "mirror or shadows" SG False intersections. Such overabundance may be disentangled by additional, individual tests for the few suspect candidates.}
\label{fig:Fig1}
 \end{center}
\end{figure}


\section{The United Group test}
\label{Sec:UnitedTest}
Let us consider as the simplest experiment, the \Purim of a sample of suspect infected ones as large as $N_{ALL}$. The simplest case might be a passenger group waiting for a flight. Let us assume, as this first simplest \Purim procedure to combine and mix all the
$N_{ALL}$ test in an unified one. This whole ($ N_{ALL}$) mixed test may be either positive to CoVid-19 or not. The absence of any infected, the negative case, will free the  $ N_{ALL}$ suspect candidates. In particular, in an airport, they may fly abroad. 
    On the contrary, if the test offer a positive result, at least one or even more components of the group must be infected. An additional inspection may be either individual (costly and time consuming)  or, as shown below,  by a quick and inexpensive  \Purim  procedure based on  partial sub-group cross-checked  tests.
    
\section{The bi-dimensional (2D) \Purim}

Let us therefore assume that the suspect group to disentangle and to analyze by \Purim may be located, one by one, within an ideal square matrix array whose components are  $N_{ALL} = N^2$. In general  $N_{ALL}$ number is not a perfect  square, but it  may  always be reduced to the sum of two (or more) comparable sub square subgroups. It may be also reduced to a quadratic group with a few  remnants candidates. This secondary procedure for an optimal fragmentation of $N_{ALL}$  into two comparable squared and to a few remnants is irrelevant here and it might be discussed elsewhere.
Anyway, one may also find an approximation  $N_{ALL} = N^2 (\pm M)$ where M is the smallest integer number needed to fill the $N^2$ square matrix. In such a case one may build  such an  square array either with $N_{ALL} = N^2$  and with a few empty (M) places, in both rows and in columns. In other cases one may arrange the patients in a larger square where:  $N_{ALL} = (N+1)^2$; consequently, as before,  few K empty places will be added in the  new arrow and new column, where:  $K = (N+1)^2- N^2 -M$. Finally any way we imagine to locate all the  $N_{ALL}$ patients into a square matrix array.
    
Therefore, as described as an example, by  Fig.\ref{fig:Fig1} ( $N_{ALL}= N^2$, $N = 9$), we may study several basic  situations. 
On the left, we assume an unique positive patient; in the center, a second with two True positive infected; the right figure shows a final case with three positive infected candidates. As shown in  Fig.\ref{fig:Fig1} a corresponding cross-check sub-group (intersection of sub-group) may disentangle immediately the unique infected one, obviously a True positive shown by a label $T$. Let us define the infected total number  by  $N_T$ as the whole number of true positive patients. Let us also define by $N_F$ the  number of false positive. 

As shown in the same figure in multiple cases arise also mirror intersection pointing to a False $F$ infected subject.
A larger number of infected samples will cause, generally, a larger number of intersection and of False $F$ infected. From the analysis of their multiple intersections there will be individuated both the infected $T$ ones but also the additional parasite false $F$ positive.
 Two False and two True in the second figure in  Fig.\ref{fig:Fig1} show the first of such situations.  Three True and five False in the last figure side show a more complex geometry. Of course the overestimation  $N_T +N_F \geq N_T$ depends on the geometrical disposal and the total number of infected in the matrix array. As obvious, in general, more True presences, more False ones. However, some rare multiple infected presences may be lead also to no False positive suspects.  We will discuss the overestimation in next section. We describe now in details the practical procedure in the  \Purim selective tests, as the few examples shown in  Fig.\ref{fig:Fig1}, to figure out the infected case by the intersection of sub-group tests.
 
 \subsection{The bi-dimensional Sub Group \Purim procedure}
    
  Imagine that all along the vertical side of the square we label each  horizontal arrow places for the candidates by a letter: a,b,c..i.  In analogy at the top of the square let us label each vertical column by the number 1,2,3--9. See  Fig.\ref{fig:Fig1}.
   Now we should call each sub group of patients $SG$ contained in an arrow or in a column, each composed by $N = 9$ tagged candidates (in column or in arrows) respectively as shown in  Fig.\ref{fig:Fig1}:
       $${SG}_a, {SG}_b,..{SG}_i$$
       $${SG}_1, {SG}_2,..{SG}_9$$
       
       The cross sections  of their cross checked test results imply, for instance, for the first case, the coexistence of two positive sub-groups:
            $$ {SG}_c, {SG}_2$$
        
            The two group  intersection identify an unique infect place subject sitting both in  ${SG}_c$ and in ${SG}_2$. He got a double positive test.
    Therefore, the patient sitting in the $c2$ matrix coordinate is the infected one.  No other  is positive nor any false suspect arise.  The  the remaining persons  are in majority  negative. Even  $77\%$ of them, are double checked.
    
     The next situation, describing $2$ positive $T$ True case, will arise by the finding of the following four subgroup positive tests:
     
            $$ {SG}_c, {SG}_2$$ 
            $$ {SG}_g, {SG}_6$$
            
      However as shown in Fig.\ref{fig:Fig1} 
      there are  also suddenly two  additional (false or "shadows") $F$ candidates, due to different "mirror" intersections:
         $$ {SG}_c, {SG}_6$$
         $$ {SG}_g, {SG}_2$$
         
        In conclusion: in present two real infected $T$ there are also two virtual false $F$ (not infected) cases. A first over-estimation of the \Purim procedures.
        In the final  third situation described by the right side in Fig.\ref{fig:Fig1}, the disposal, with the three True positive candidate induced  a total of 6 false positive candidates. In general the suspects  (real or True ones $N_T$) are less than the suspected  ones $P_s \geq N_T+ N_F$ ones.

    The last case in the  Fig. \ref{fig:Fig1}  shows  the eventual three True positions and  the presence of six suspected  positive tests; the True infected ones are located by the sub-group intersection below:
            $$ {SG}_c, {SG}_2$$
            $$ {SG}_g, {SG}_6$$
            $$ {SG}_e, {SG}_8$$.
             
       However, there are additional 6 independent sub-group intersections  defining six more possible or virtual positive, False, candidates:
             
            $$ {SG}_c, {SG}_6$$
            $$ {SG}_g, {SG}_2$$
            $$ {SG}_e, {SG}_2$$.
            $$ {SG}_e, {SG}_6$$
            $$ {SG}_c, {SG}_8$$
            $$ {SG}_g, {SG}_8$$.
            
            All of these 9 possible intersections or infected cases must be tested individually or by similar mixed and cross-checked  procedure. Therefore, the \Purim procedure has a quite good accuracy for diluted sample of infect cases.
            Naturally  the present pandemia is, hopefully,  below $1\%$ of the population and in such a group (nearly hundred) sample, as in Fig. \ref{fig:Fig1}, \Purim procedure will detect positive subjects with quite a good precision. Therefore,  we may already conclude that in general the  single tests versus the  \Purim procedure  does cost, in economy and in time, by a ratio $R= N^2/(2N)$. In a hundred sample, a \Purim procedure may be 5  times more inexpensive and faster than individual tests. In a thousand sample number (located in a square array $N$ of  nearly $32\cdot 32$ size) the cost in time and in expenses for the ideal (just one positive) \Purim case would be only $6.4\%$ of $N^2$ tests;  if there are a few True infected, as $<5$ candidates or about $0.5\%$, the additional mirror or false candidates  would  lead  to a final total $8.9\%$ of the cost respect to  $N^2 = 1024$ single tests expense. The whole saving will be above $91\%$.
        
\begin{figure}[!htb]
\begin{center}
\hfill
\includegraphics[width=0.46\columnwidth]{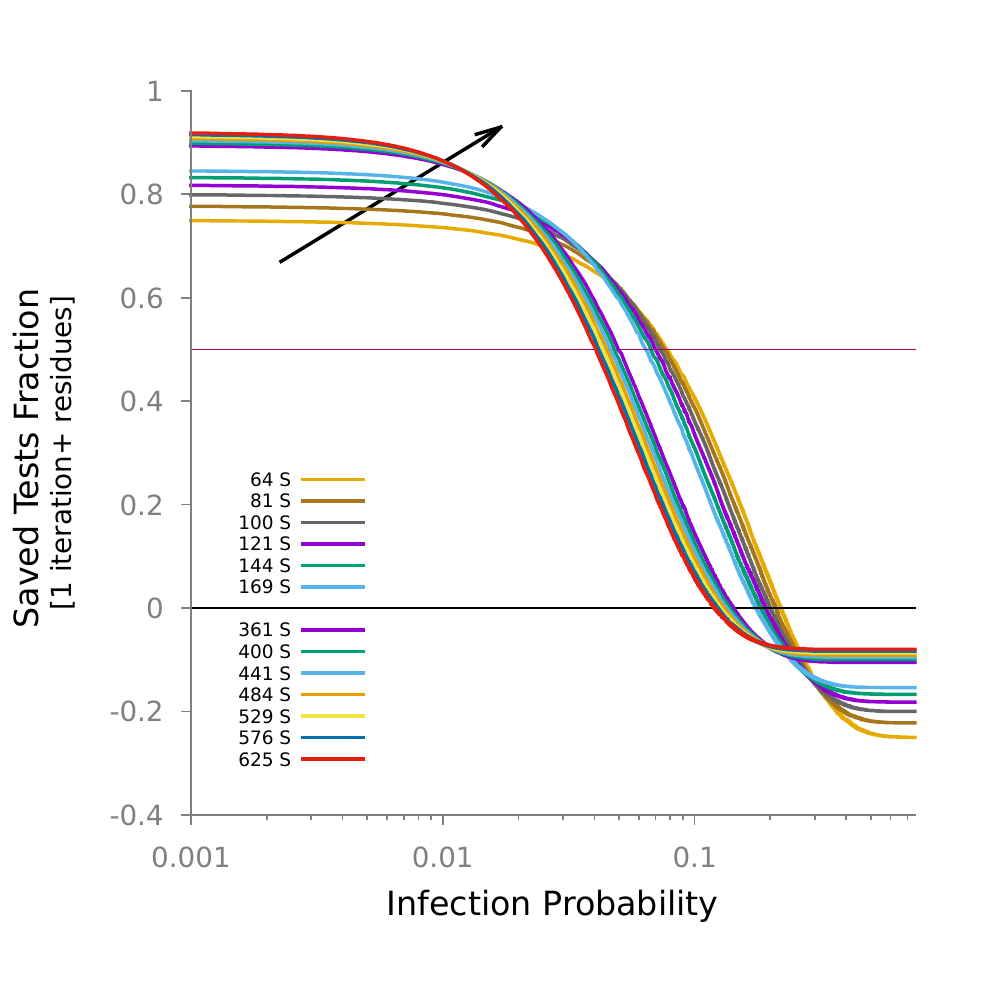}\hfill
\includegraphics[width=0.46\columnwidth]{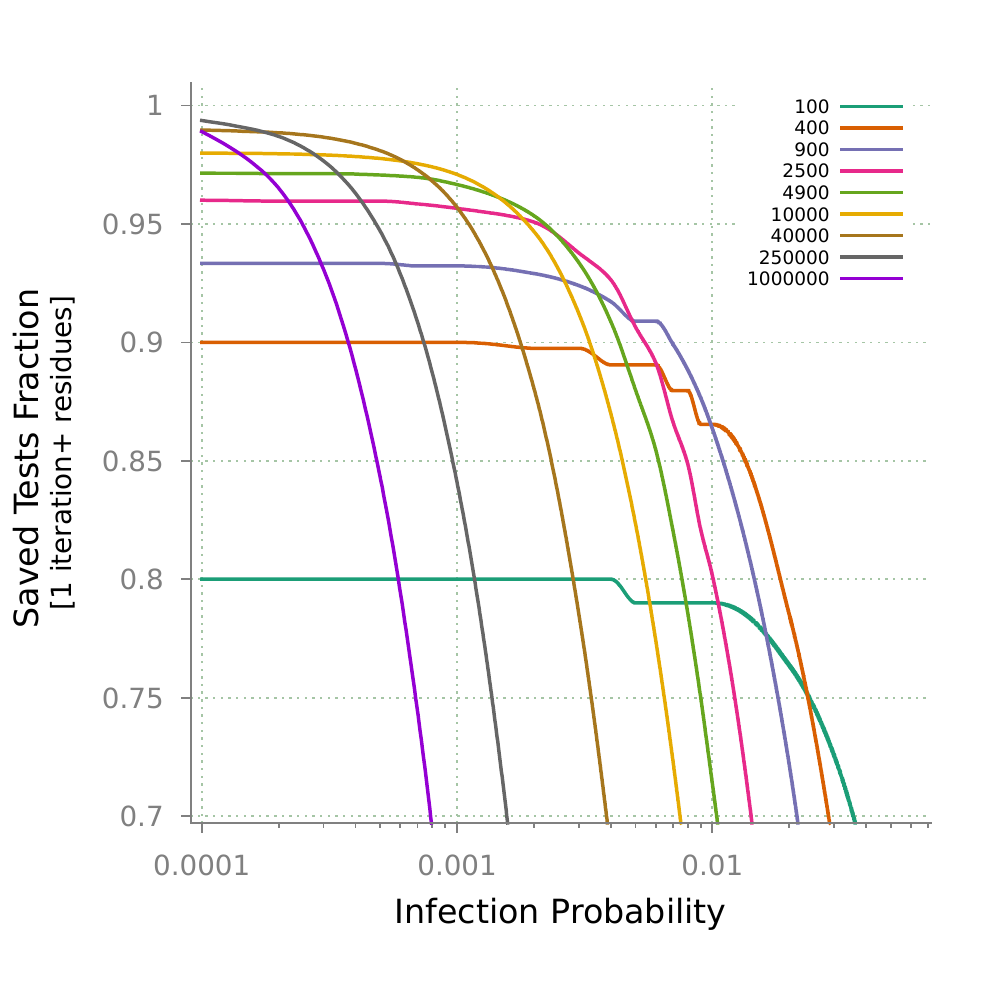}
\hfill
\caption{\textsc{Saved Tests} The figures show the trend with respect to the probability of infection $P_i$ of the fraction of tests that is saved with the \Purim procedure. On the left the overall trend, on the right a zoom on a specific area. The usual procedure involves a single test for each individual: if the individuals are $N_{All}$, then $N_{All}$ tests are performed. The \Purim procedure, as explained in the text, provides for the clustering of individual tests in groups with N $= {N_{All}}^{1/2}$ individuals each. It also prescribes to put the sample of each individual in $D$ (here equal to two) groups with the additional condition that each pair of individuals can be present simultaneously in only one group.  We define $T_s$ the number of tests saved as $T_s = N^2 -2N -N_p$,  $N_p$ is the average number of individuals who tested positive. The figure shows, by a MonteCarlo the best trend of the saved test fraction $T_s / N_{All}$ as a function of the probability of infection $P_i$ and for different values of $N_{All}$. The figure point out  that the best efficency is obtained with a number of subgroups $N$ which depends in a good approximation (inversely) on the probability of the infection.}
\label{fig:FigSavedTests}
 \end{center}
\end{figure}
        
\begin{figure}[!htb]
\begin{center}
\includegraphics[width=0.46\columnwidth]{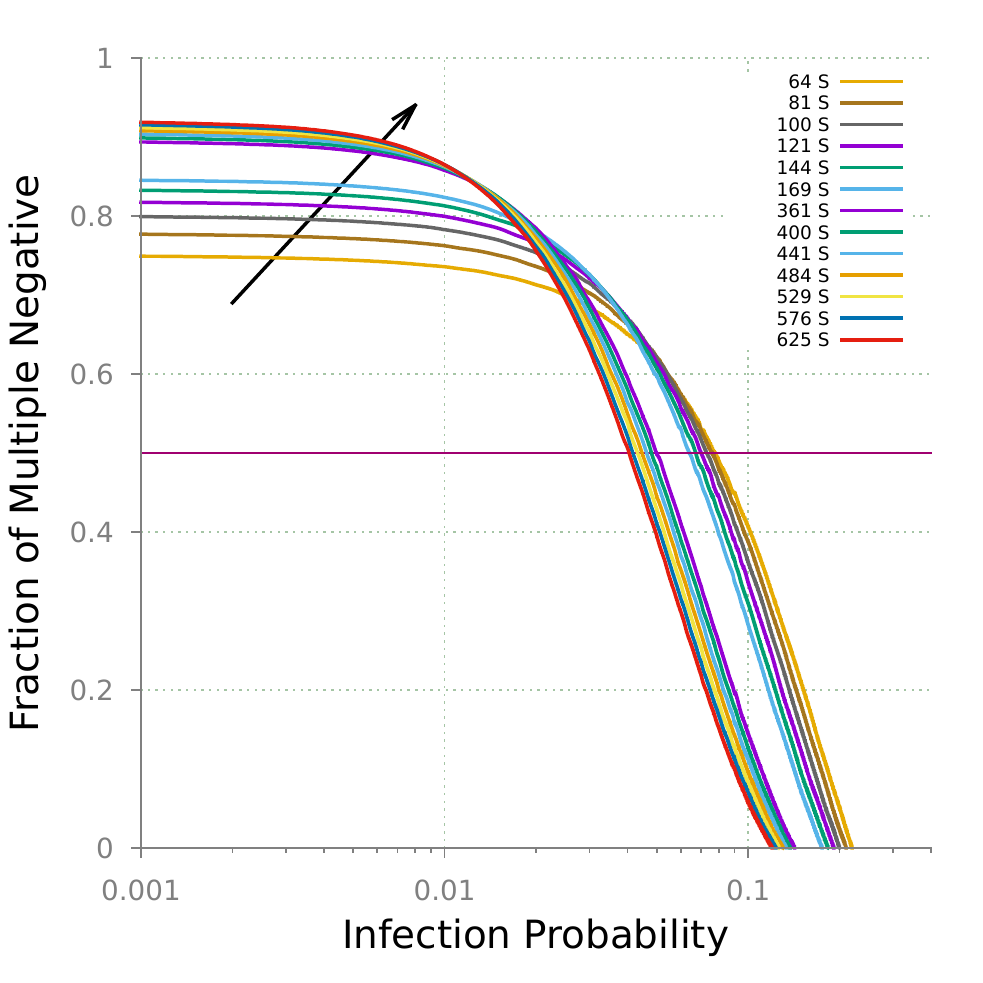}\qquad
\includegraphics[width=0.46\columnwidth]{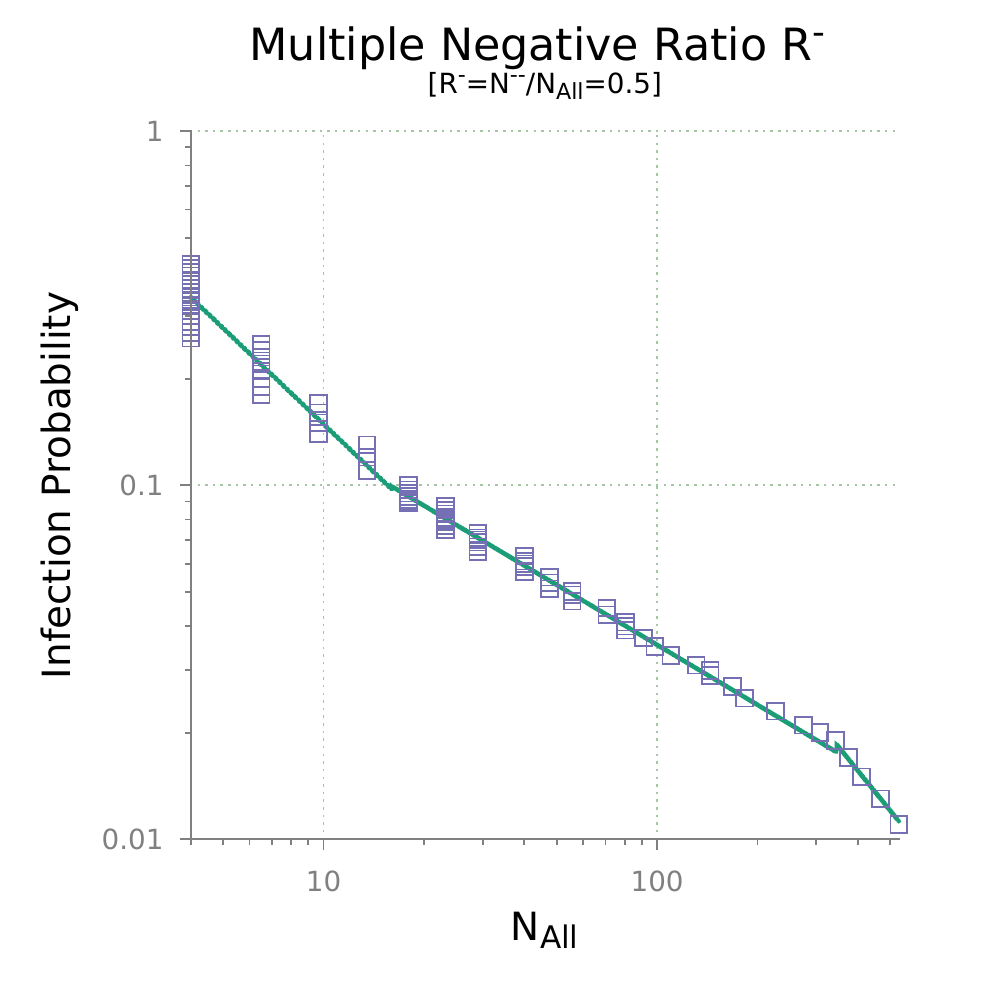}
\caption{\textsc{Multiple Negative Tests.} From the \Purim procedure some of the sample can result negative in all subgroup tests, hence multiple negatives, $N^{- -}$. In the left pane, the curves for different values of $N_{All}$, the subgroups dimensions growing with the arrow,  with the ratio $N^{- -}/N_{All}$, as a function of the probability of infection $p$. In the right panel the intersections of these curves with the value of 0.5.}
\label{fig:FigMultipleNegatives}
 \end{center}
\end{figure}
                
\subsection{Over-estimation in a crowded sample}             
\begin{figure}[!htb]
\begin{center}
\includegraphics[width=0.45\columnwidth]{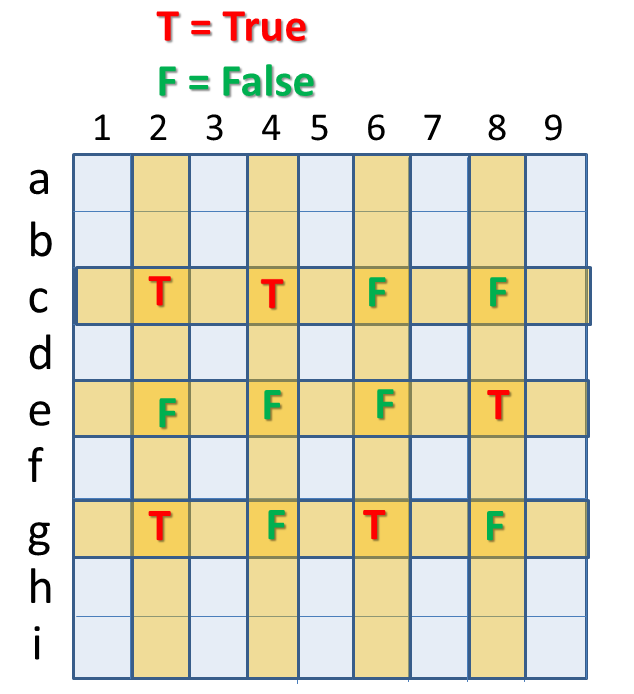}
\caption{\textsc{CrossCheck} As in the previous Fig.\ref{fig:Fig1} multiple intersections may occur simultaneously, leading to a complex situation with the presence of false positives. In this case with 5 True and 7 False positives. This redundancy is limited and can be disentangled by further tests, only on the 12 positives.}
\label{fig:Fig1B}
 \end{center}
\end{figure}

\begin{figure}[!htb]
\begin{center}
\includegraphics[width=0.90\columnwidth]{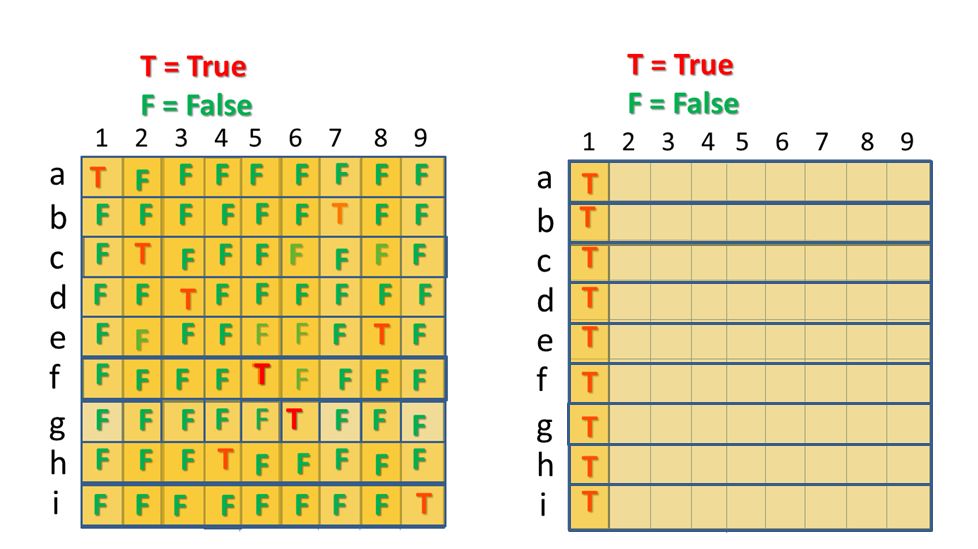}

\caption{\textsc{Opposite configurations} Two extreme fine tuned and  antithetic configurations may be able to affect the outcome of the \Purim procedure in an opposite way.  On the left side the N infected samples are each in a different row or column. It is the equivalent of having them on the diagonal. The test result will not discriminate, indicating all $N^2$ subjects as possible positive. In the right part instead, the N infected are all positioned on the same column (or row), for instance the first subgroup ${SG}_1$. The result is exhaustive by identifying all N True positive subjects without any False positive.}
\label{fig:Fig1C}
 \end{center}
\end{figure}

We may wonder about the previous overestimation in \Purim procedure for a crowded source positive presence.
There might be not always  a simple quadratic growth of the  arrow and column intersections, as in the case described in Fig. \ref{fig:Fig1B}.
   The general  complexity is not obvious.
   For very fine tuned locations of $N_{T}=N$ infected subjects, each of them with a different number of row and column, they may produce, $2\cdot N$, all columns and rows positive, implying  $N^2$ cross-checked intersections, suggesting that all of the sample is infected. Indeed, in the case described below only N are True positive while  $N^2-N$ are just False positive; see for instance left side in Fig. \ref{fig:Fig1C}. Such a rare situation may occur once the sample of infected has a large probability P to be present within the total group $$P \geq \frac{1}{ 4\cdot {N_{ALL}}^{1/2}}$$. For a square of 32 element and ${N_{ALL} = 1024}$, this occur for nearly a large infected presence $\geq 0.75\%$, or 7 or 8 infected patients. An extreme additional,  fine tuned case occurs when only any arrow of the sub-group (in the example, the first vertical one) is identical positive for all sub-groups while all the others (horizontal) sub-groups are different ones: in that "well aligned" configuration, both horizontal or vertical, there are no False, but just N True cases as shown in the right side in Fig. \ref{fig:Fig1C}.

 In general for a number of $N_H$ (horizontal) positive row and for a number of $N_V$ (vertical) column sub-group positives, the resulting number of True $N_T$ cases is bounded by a maximal and minimal value:
\begin{equation}
   (Max (N_H, N_V)) \leq N_T \leq (N_H\cdot N_V) .
\end{equation} 
In analogy the number of false negatives $N_{F}$ is analogously bounded by
\begin{equation}
   (0 \leq N_F \leq (N_H\cdot N_V)- (Max(N_H, N_V))).
\end{equation}

 For example in the simple case as in Fig. \ref{fig:Fig1B} where $N_H =3$, $N_V =4$ the previous bounds for  $N_T$ becomes:$  ( 4 \leq  N_T \leq 12)$. Fig. \ref{fig:Fig1B} shows indeed  $N_T = 5$ True and $N_F = 7$ as in their bounded limits.
                
   Naturally, because $N_H$ and  $N_V$ are in general much less N, and if  $N^2$ is a large number (hundreds or above) and the present false cases are rare (a few unities). Therefore additional test are not too costly.
    There are some  cases where the True sample may be aligned in an axis or in all diagonal places;  rare situations where False will be ruling or will be absent at all: see  Fig.\ref{fig:Fig1C}. These situations might be over estimating because the infect presence is not negligible ($N$ versus $N^2$).
            
The bidimensional cross-cheked sub-group procedure might be extended to multi-dimensional \Purim procedure, as it is  shown in a more rich but more complex correlations approach considered below.

\section{The Multi-Dimensional \Purim procedure}   

The natural extensions of the previous bi-dimensional \Purim procedure could be described for geometrical visibility by a cubic array where the sample to be tested would be located each one to a mini-cell tagged by its 3D coordinate.The cubic matrix may be imagined as in figure Fig. \ref{fig:Fig2}. The candidates are located in $3D$ cell structures labeled by three array axis. Two quite different  different \Purim procedure may used to disentangle the positive infect subjects; a planar one or a vector one, base on different
    cross-checked procedure.

\subsection{The Planar (2D) cross-check inside a 3D array }
\begin{figure}[!htb]
\begin{center}
\includegraphics[width=0.95\columnwidth]{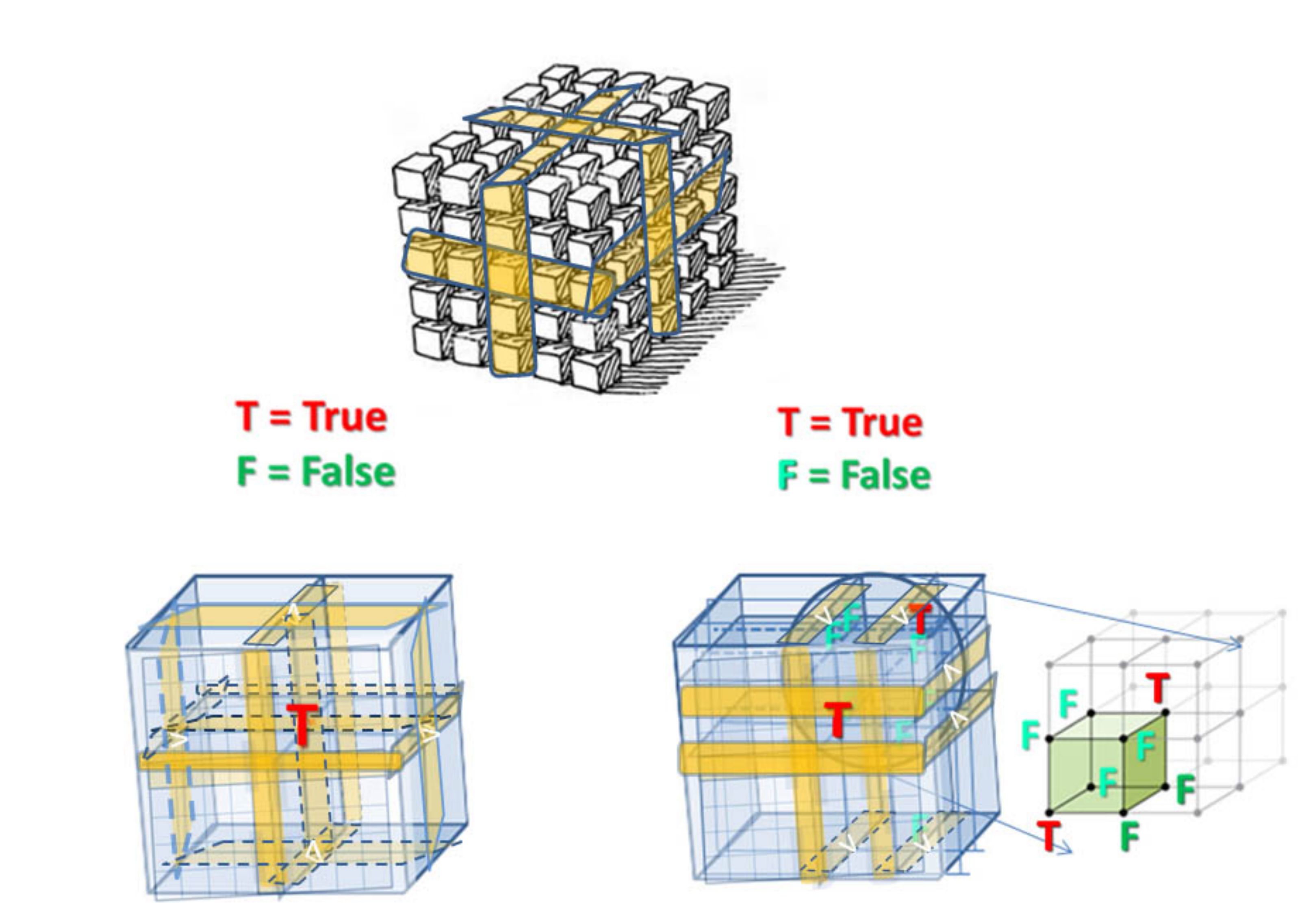}
\caption{\textsc{} The 3D array distribution of the candidates  in a cell array $6^3$  tagged by horizontal, Vertical and Longitudinal labels. For sake of simplicity in upward figure, the array cells are  shown  with the colored plane sections that are crossing and selecting an inner central True infected case. The  inner detection  by the independent 3 sub-planar-groups is shown on the left side. Moreover, as shown in the right side, the presence of just 2 True infect cases creates six plane intersection forming  $2^3$ cubic vertex: $2 N_T + 6 N_F = 8$. Therefore  a larger growth of False suspected cases. The geometry is zoomed in the right corner to show the cubic structure. This False growth imply a wider over-estimation respect to the bi-dimensional cases. } 
\label{fig:Fig2}
 \end{center}
\end{figure}

Let us imagine to locate the $N_{ALL}$ suspected patients   inside a cube cell defined by its 3D coordinates. As in the bi-dimensional case, this may be reached by dividing the whole number $N_{ALL}$ in comparable sub cubic numbers or by  to located the surplus candidate into few empty cells in such a larger cubic array location. The different alternative to optimize $N_{ALL}$ into cubic sizes are irrelevant here.
  Each person is tagged by its $H_i$, horizontal,  $V_i$, vertical,  $L_i$, longitudinal position inside the  ideal cubic cell volume. For instance a sample of  $ N_{ALL} = N^3 = 216$  passengers are tagged by a 3D array made by $6^3$ cubic side cell structure as shown in Fig. \ref{fig:Fig2}. The easy way to act within a planar  \Purim procedure in 3D cells (candidates) array is based on  the construction of $3\cdot N$ planar sub-groups associated to all the three possible planes: $HV_i$, $HL_j$, $LV_k$. (i,j,k are bounded within N cube size). Each of the planar array is made by $N^2$ sub-group components united by their test. In present example each planar sub-group contains $N^2= 36$ patients tests united. These sub groups ($3 \cdot 6$) are organized and tested independently.  Therefore one needs only a total of $3\cdot6 = 18$ test to intersect to discover potential (True and False) infected cases and not  $ N_{ALL} = N^3 = 216$.  In the presence of an unique positive person in the whole cube sample its planar intersections may define just a single cell candidate,  as shown in Fig. \ref{fig:Fig2} left side correlated to the upper figure.
 Such a unique  identification (one True) is  well defined.
In the very diluted sample this procedure is extremely effective and economic. The ratio between  $3\cdot N$  and $N^3$ test is $3/N^2$. Therefore very inexpensive.
     In the present $ N_{ALL}= 216$ case the cost is below $10\%$.
     There are of course mirror effects due to many parasite $F$ False intersections, even more than in previous  bi-dimensional case. As shown in the zoomed side in Fig. \ref{fig:Fig2} the 2 True  positive cell location, defining a complete inner mini-cube, they produce 6 additional planar intersections whose corner are forming 6 False or virtual infected cases by the other two $T$ True infected presence.
     Indeed the total intersections of  three independent planar sub-group  are the exactly the corners of a cube, that are in a total of $8$. 

 The number of 3D planar intersections are very often leading to an  over-estimation of True candidates whose number may be nevertheless well bounded.
   For instance a simple generalization may occur for any three True location, all with different coordinates. Their tests will tag $3\cdot 3$ positive planes  results. The consequent whole possible intersections will be much larger than the few three of the True cases: indeed the $3^3$ planar intersections will suggest (at each corner), $27$ suspected positive cell candidates. In the suggested case, assuming non co-planar $3$ True patient, there will be  $3^3- 3$ False ones. A very large polluted result.
       Naturally there may be also some coexistence, in the same plane of few True positive subject, reducing the False overabundance. The complexity may be restricted to the extreme bounds for the $N_T$ ones.
Let us assume that there are $N_H$ horizontal positive planes, $N_V$ vertical and $N_L$ longitudinal positive planes. Their total vertex (or intersections) number will be: $N_{Vertex}= (N_H \cdot N_V \cdot N_L)$. There might be in general a True number $N_T$ inside bounded by:

\begin{equation}
   (Max (N_H, N_V, N_L)) \leq N_T \leq (N_H\cdot N_V \cdot N_L) .
\end{equation} 

In analogy the False $N_{F}$  are also bounded by
\begin{equation}
   (0 \leq N_F \leq (N_H\cdot N_V \cdot N_L)- (Max(N_H, N_V,N_L))).
\end{equation}

In the Fig. \ref{fig:Fig2} these bounds are confirmed. 
The true events are $N_T =2 $, the False are $N_F = 6 = 8-2$. Both are following their extreme  bound limits.

\begin{figure}[!htb]
\begin{center}
\includegraphics[width=0.95\columnwidth]{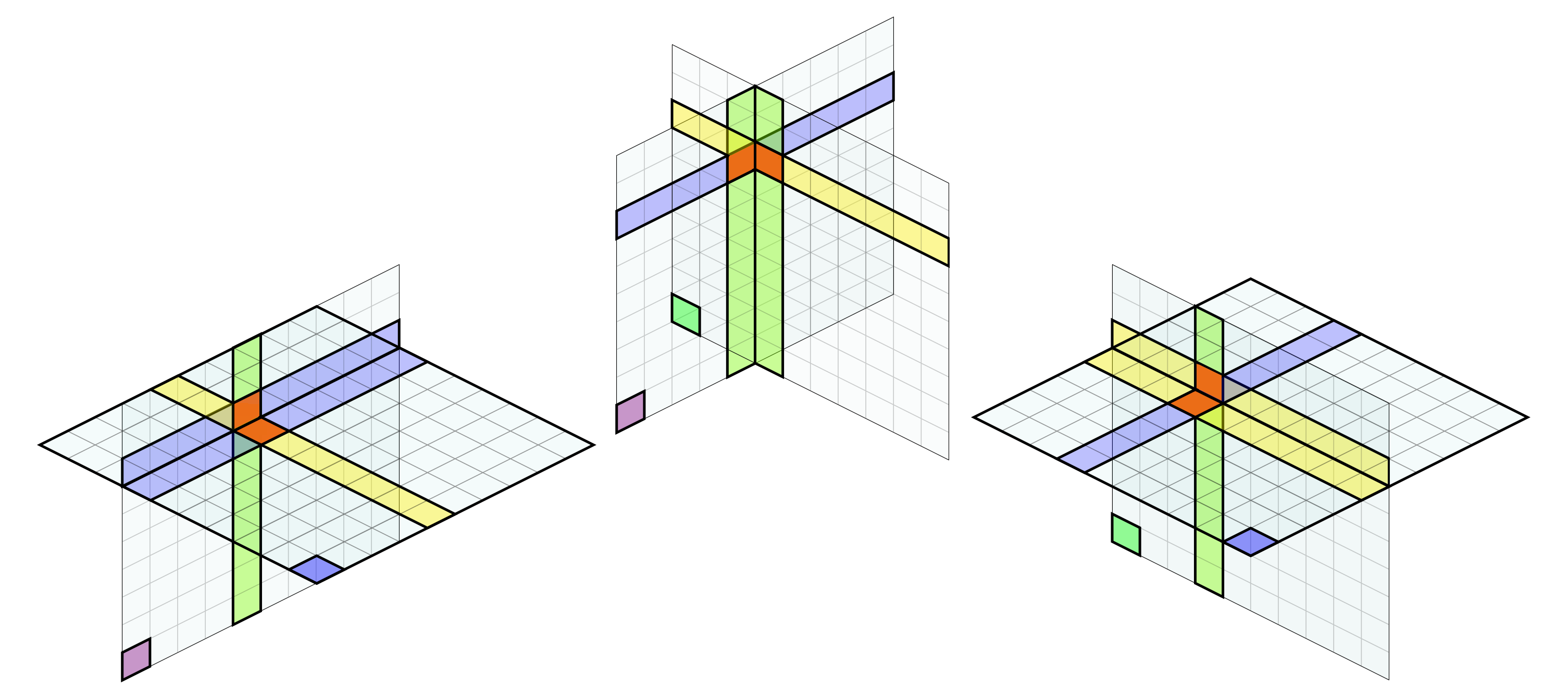}
\caption{\textsc The candidates are tagged on the horizontal, vertical and longitudinal axis. The construction of independent sub-vector-groups by their 3D intersection, may identify the True positive infect case with redundancy. This vector procedure requires many more tests $3 N^2$ and cost but it imply less  overestimation of False events respect to the previous planar procedure.} 
\label{fig:Fig2-Idea}
 \end{center}
\end{figure}

\subsection{The Vector (1D) cross-check inside a cube array }

 The gain by planar inspection is obvious:  to collect larger planar  sub-group number  is a cheaper way to inspect a very large sample. However the cost is paid in the over-estimation (for polluted overcrowded positive populations). The need for additional tests is reducing the planar \Purim advantage.
    There  is an additional tool in the \Purim procedure based on the vector (column) sub-group in one dimension section. In this case the cross-check is in general overabundant and redundant: this procedure may avoid many False presences. However, this accuracy is paid by the growth of test vector needed tests: see Fig.\ref{fig:Fig2-Idea}. Assuming as above $N_{ALL}= N^3$ one needs $N^2$  column or vector sub groups, each one of $N$ cell or patients, for each cube face side. This imply $3\cdot N^2$ tests respect to the individual $N^3$. For large numbers this approach may be interesting, but the gain or   the huge cost-saving (the ratio $R_{gain}$ between 1 dimension procedure and individual one)  is only $R_{gain1} = 3 \cdot N^2/ (N^3) = 3/N$, while in previous one (for the 2D array structure) it was  $R_{gain_2} = 3/(N^2)$. For $N=10$ and a thousand of the whole candidates, this imply for the planar \Purim just $R_{gain_2} = 3\%$ while for a vector 1 dimensional procedure a less attractive: $R_{gain1} = 30\%$.

\section{The Multi-Dimensional (3D,..MD)  \Purim}
 The generalization to  more  M dimension is obvious: the same extension may be done from the third dimension case to more, M, dimensions. The procedure is less intuitive  but it is very simple, not a necessary step to discuss here in detail. The chain of sub structure (planar, vector) for \Purim procedures might be extended to (Cubic, planar, vector) and so on for each greater dimension stage. The most effective procedure may have an ratio R (among \Purim test versus single ones) as $R= M \cdot  N/(N^3)= M/(N^{M-1}) = M/({{N_{ALL}}}^ {(M-1)/M}) $. Therefore the larger M and the better the effective R (assuming a very diluted sample of candidates). 
 
 \subsection{The zoom scaled   \Purim  filter}
   The procedure  that we are suggesting might be seen  as a filter  able to pick up, by its narrow orthogonal array net, the single or just a few positive infect patients.
     The same filter may applied in larger scale.
      Indeed the \Purim procedure may be enlarged in wider scale changing the granularity of the sample. Let us imagine a whole clustered group (hundreds or thousands persons), whose mixed test are jointed and unified, as a single subject, whose collective test is either a positive or negative one. Such a positive clustered group would contain at least an infected person. Now, let us imagine a larger scale array, made by such clustered groups, each individually located in a square matrix, of such clustered groups. Each of those sub-group as a whole may be tagged as positive or negative. For example, let us imagine that the clustered groups are  the school classes of $25$ students each. Let us also suggest, that the total class number in the big school is: ( $Ncl_{ALL}$ =25).
   Therefore we have a $625$ total students in the whole school. The classes are identify, for instance, by their age ($14th-18th$ year old) and also by their peculiar section (a,b,c,d,e), into a wider square matrix classes-elements. 
   Each class may be localized by its letter and class age. Such an array $5x5$ of clustered groups (= classes) resemble, in full analogy,  the previous bi-dimensional \Purim section. We may find by column and row  intersection the (one or few) infected (true or false) positive classes, with just 10 unified tests. We verify the $25$ classes. If any is positive, Finally we may further inquire,  restricting as in a microscopic zoom,  the positive class, into a narrow scale tests, looking for the positive and negative or false individual students.
   Any positive class will be at once again re-analyzed, by the inner zoomed \Purim procedure, to figure out each individual infected student.   With  only a first $10$ tests we may inspect $625$ student classes.
     Like from a telescope toward a  microscope view, from the large  to the small scale, \Purim procedure may  offer rapid, inexpensive, as well a high filter accuracy.

\section{Discussion}

The proposed method is suitable for detecting suspects, but also identifies as possible positive subjects that would have been clinically negative due to a "shadow effect" (F). As far as the method is used for screening large groups with low probability of being infected, this reduction of accuracy won't decrease its ability to avoid the spreading of the infection.

The effectiveness of the method in comparison to one-by-one testing is related to the increase in the number of people tested together as a unique sample. Yet, this is limited by the probability of each individual being infected, as shown in Fig.~\ref{fig:FigSavedTests}. 
Another limit is due to the dilution of the biological sample by mixing of the positive ones with the negative ones in the same sub-group. This limit depends on both the concentration of the nCoV in the samples collected from infected people and on the limit of detection (LoD) of the testing technology. 
No specific data on the actual LoD is available in the open literature. The CDC reports \cite{fda2019-Covid} a limit for common RT-PCR lies between $10^{0.0}$ and $10^{0.5}$ RNA copies $\mu L$. Yet no data are available on the concentration of RNA copies in the specimens of infected people. On the other hand, authors from China \cite{Xie2020,Zu2020} report relevant negative PCR tests in symptomatic patients, even with repeated tests. Nonetheless, a dilution between 10 to 100 would be probably acceptable leading to the opportunity for rapid testing of clusters between 100 and 10,000 people with reduction of number of laboratory tests between $80\%$ and $98\%$.
In a limited time scenario, in which the number of people tested is limited by the availability of laboratory time, this means that the application of the proposed method will allow to increase the number of tested people between 5 times and 50 times in comparison to standard one-by-one testing.

\section{Applications and Conclusions}
 The present common project to defeat the CoVid-19 by a severe quarantine of the whole national  economy might be successful, but extremely costly and late.
   The ability to distinguish and filtering positive from negative subjects may free a steady component of the population into the economy life, while holding the rest of the positive or untested citizens in  a restrictive  quarantine.
The usual one-test to each patient in present  international CoVid-19 pandemia is exhausting both in the economy and in time prospective. A fast and economic  is based on the \Purim procedure (\PurimLong): the collective array sub-group tests whose cross-checked results may rapidly intersect and disentangle the real and the few suspect positive patients.    
  For a flight or a ship embark procedure, respectively of  $10^2-10^4$ individuals passengers the present \Purim strategy may cost about ($20\%$) up to ($2\%$) of the whole sample test, both in cost and in time.
 In a diluted $<1 \%$  infected population stages  these procedure might require only a negligible additional tests to better identify the real  True positive infected passengers, from the few possible False ones. In general such \Purim test will free and guarantee the majority  of the passengers as being negative  by a double-checked sub-group test. The \Purim  application in the present medical subjects it may avoid the presence of infected doctors at the hospital care.  The same \Purim purification may be applied to classes or to schools leading,  step by step, to a fast reopening of the education system. For an ideal  national  population, as large as the Italian one of $ 6\cdot 10^7$ candidates, the fragmentation of the whole population into $6\cdot 10^3$ sub group, each made by  $N_{ALL} =10^4$ citizens, may be achieved by $2 \cdot N^2$, just 2 hundreds, sub test each, within  a realistic goal: minor city  areas may be tested step by step and they may be let  free in  life while positive identified subjects may be better guaranteed in their medical and quarantine needs.  This procedure would cost nearly $2-3 \%$ of the individual national population test, with possibly dozens of billions of euro saving expenses. The procedure might be naturally extended in the international arena,  where the pandemia is just at the earliest stages, with more profit and success.


\section*{Acknowledgement}
The authors wish to thank the Prof. Moshe Labi, of New York A. Einstein Hospital, for very helpfull suggestions and for letting us know today of a few days ago  Jerusalem Post~\cite{Jean.2020} article, regarding a Technion proposal,  analogous to our first and simplest United Group \Purim procedure [see section \ref{Sec:UnitedTest}]. 

\section*{References}
\nocite{*}
\bibliography{QuickDetectionMethodBIB}

\begin{thebibliography}{10}
\expandafter\ifx\csname url\endcsname\relax
  \def\url#1{\texttt{#1}}\fi
\expandafter\ifx\csname urlprefix\endcsname\relax\def\urlprefix{URL }\fi
\expandafter\ifx\csname href\endcsname\relax
  \def\href#1#2{#2} \def\path#1{#1}\fi

\bibitem{huang2020clinical}
C.~Huang, Y.~Wang, X.~Li, L.~Ren, J.~Zhao, Y.~Hu, L.~Zhang, G.~Fan, J.~Xu,
  X.~Gu, et~al., Clinical features of patients infected with 2019 novel
  coronavirus in wuhan, china, The Lancet 395~(10223) (2020) 497--506.

\bibitem{bianconi2020oswald}
A.~Bianconi, A.~Marcelli, G.~Campi, A.~Perali, Oswald growth rate in controlled
  covid-19 epidemic spreading as in arrested growth in quantum complex matter
  (2020).
\newblock \href {http://arxiv.org/abs/2003.08868} {\path{arXiv:2003.08868}}.

\bibitem{cinelli2020covid}
M.~Cinelli, W.~Quattrociocchi, A.~Galeazzi, C.~M. Valensise, E.~Brugnoli, A.~L.
  Schmidt, P.~Zola, F.~Zollo, A.~Scala, The covid-19 social media infodemic,
  arXiv preprint arXiv:2003.05004.

\bibitem{Gralinski2020}
L.~E. Gralinski, V.~D. Menachery,
  \href{https://doi.org/10.3390/v12020135}{Return of the coronavirus:
  2019-{nCoV}}, Viruses 12~(2) (2020) 135.
\newblock \href {http://dx.doi.org/10.3390/v12020135}
  {\path{doi:10.3390/v12020135}}.
\newline\urlprefix\url{https://doi.org/10.3390/v12020135}

\bibitem{doi:10.1056/NEJMoa2001017}
N.~Zhu, D.~Zhang, W.~Wang, X.~Li, B.~Yang, J.~Song, X.~Zhao, B.~Huang, W.~Shi,
  R.~Lu, P.~Niu, F.~Zhan, X.~Ma, D.~Wang, W.~Xu, G.~Wu, G.~F. Gao, W.~Tan, A
  novel coronavirus from patients with pneumonia in china, 2019, New England
  Journal of Medicine 382~(8) (2020) 727--733.
\newblock \href {http://dx.doi.org/10.1056/NEJMoa2001017}
  {\path{doi:10.1056/NEJMoa2001017}}.

\bibitem{Zhou_2020}
T.~Zhou, Q.~Liu, Z.~Yang, J.~Liao, K.~Yang, W.~Bai, X.~Lu, W.~Zhang,
  \href{http://dx.doi.org/10.1111/jebm.12376}{Preliminary prediction of the
  basic reproduction number of the wuhan novel coronavirus 2019‐ncov},
  Journal of Evidence-Based Medicine 13~(1) (2020) 3–7.
\newblock \href {http://dx.doi.org/10.1111/jebm.12376}
  {\path{doi:10.1111/jebm.12376}}.
\newline\urlprefix\url{http://dx.doi.org/10.1111/jebm.12376}

\bibitem{Araujo2020.03.12.20034728}
M.~B. Araujo, B.~Naimi,
  \href{https://www.medrxiv.org/content/early/2020/03/16/2020.03.12.20034728}{Spread
  of sars-cov-2 coronavirus likely to be constrained by climate}, medRxiv\href
  {http://arxiv.org/abs/https://www.medrxiv.org/content/early/2020/03/16/2020.03.12.20034728.full.pdf}
  {\path{arXiv:https://www.medrxiv.org/content/early/2020/03/16/2020.03.12.20034728.full.pdf}},
  \href {http://dx.doi.org/10.1101/2020.03.12.20034728}
  {\path{doi:10.1101/2020.03.12.20034728}}.
\newline\urlprefix\url{https://www.medrxiv.org/content/early/2020/03/16/2020.03.12.20034728}

\bibitem{to2020consistent}
K.~K.-W. To, O.~T.-Y. Tsang, C.~C.-Y. Yip, K.-H. Chan, T.-C. Wu, J.~M.-C. Chan,
  W.-S. Leung, T.~S.-H. Chik, C.~Y.-C. Choi, D.~H. Kandamby, et~al., Consistent
  detection of 2019 novel coronavirus in saliva, Clinical Infectious Diseases.

\bibitem{fargion2014howwhen}
B.~I. Fargion, How,when and how much a card deck is well shuffled? (2014).
\newblock \href {http://arxiv.org/abs/1407.6950} {\path{arXiv:1407.6950}}.

\bibitem{fda2019-Covid}
Cdc 2019-novel coronavirus (2019-ncov) real time rt-pcr diagnostic panel,
  document cdc-006-00019, revision: 02,
  \url{www.fda.gov/media/134922/download}, accessed: 2019-03-25.

\bibitem{Xie2020}
X.~Xie, Z.~Zhong, W.~Zhao, C.~Zheng, F.~Wang, J.~Liu,
  \href{https://doi.org/10.1148/radiol.2020200343}{Chest {CT} for typical
  2019-{nCoV} pneumonia: Relationship to negative {RT}-{PCR} testing},
  Radiology (2020) 200343\href {http://dx.doi.org/10.1148/radiol.2020200343}
  {\path{doi:10.1148/radiol.2020200343}}.
\newline\urlprefix\url{https://doi.org/10.1148/radiol.2020200343}

\bibitem{Zu2020}
Z.~Y. Zu, M.~D. Jiang, P.~P. Xu, W.~Chen, Q.~Q. Ni, G.~M. Lu, L.~J. Zhang,
  \href{https://doi.org/10.1148/radiol.2020200490}{Coronavirus disease 2019
  ({COVID}-19): A perspective from china}, Radiology (2020) 200490\href
  {http://dx.doi.org/10.1148/radiol.2020200490}
  {\path{doi:10.1148/radiol.2020200490}}.
\newline\urlprefix\url{https://doi.org/10.1148/radiol.2020200490}

\bibitem{Jean.2020}
C.~Jean,
  \href{https://www.jpost.com/HEALTH-SCIENCE/Acceleration-in-multiple-coronavirus-tests-at-once-by-Israel-research-team-621533}{Israeli
  team has coronavirus test kit to test dozens of people at once}, Jerusalem
  Post.
\newline\urlprefix\url{https://www.jpost.com/HEALTH-SCIENCE/Acceleration-in-multiple-coronavirus-tests-at-once-by-Israel-research-team-621533}

\bibitem{poon2003early}
L.~L. Poon, K.~H. Chan, O.~K. Wong, W.~C. Yam, K.~Y. Yuen, Y.~Guan, Y.~D. Lo,
  J.~S. Peiris, Early diagnosis of sars coronavirus infection by real time
  rt-pcr, Journal of Clinical Virology 28~(3) (2003) 233--238.

\end{thebibliography}

\end{document}